\documentclass[useAMS,usegraphicx,usenatbib]{mn2e}

\usepackage{times}
\usepackage{amssymb}
\usepackage{amsmath}

\begin{document}

\include{defn}
\def\cm{{\rm\thinspace cm}}
\def\gm{{\rm\thinspace gm}}
\def\dyn{{\rm\thinspace dyn}}
\def\erg{{\rm\thinspace erg}}
\def\eV{{\rm\thinspace eV}}
\def\MeV{{\rm\thinspace MeV}}
\def\g{{\rm\thinspace g}}
\def\ga{{\rm\thinspace gauss}}
\def\K{{\rm\thinspace K}}
\def\keV{{\rm\thinspace keV}}
\def\km{{\rm\thinspace km}}
\def\kpc{{\rm\thinspace kpc}}
\def\Lsun{\hbox{$\rm\thinspace L_{\odot}$}}
\def\m{{\rm\thinspace m}}
\def\Mpc{{\rm\thinspace Mpc}}
\def\Msun{\hbox{$\rm\thinspace M_{\odot}$}}
\def\Zsun{\hbox{$\rm\thinspace Z_{\odot}$}}
\def\pc{{\rm\thinspace pc}}
\def\ph{{\rm\thinspace ph}}
\def\s{{\rm\thinspace s}}
\def\yr{{\rm\thinspace yr}}
\def\sr{{\rm\thinspace sr}}
\def\Hz{{\rm\thinspace Hz}}
\def\MHz{{\rm\thinspace MHz}}
\def\GHz{{\rm\thinspace GHz}}
\def\chisq{\hbox{$\chi^2$}}
\def\delchi{\hbox{$\Delta\chi$}}
\def\cmps{\hbox{$\cm\s^{-1}\,$}}
\def\cmpssq{\hbox{$\cm\s^{-2}\,$}}
\def\cmsq{\hbox{$\cm^2\,$}}
\def\cmcu{\hbox{$\cm^3\,$}}
\def\pcmcu{\hbox{$\cm^{-3}\,$}}
\def\pcmcuK{\hbox{$\cm^{-3}\K\,$}}
\def\dynpcmsq{\hbox{$\dyn\cm^{-2}\,$}}
\def\ergcmcups{\hbox{$\erg\cm^3\ps\,$}}
\def\ergpcmps{\hbox{$\erg\cm^{-3}\s^{-1}\,$}}
\def\ergpcmsqps{\hbox{$\erg\cm^{-2}\s^{-1}\,$}}
\def\ergpcmsqpspA{\hbox{$\erg\cm^{-2}\s^{-1}$\AA$^{-1}\,$}}
\def\ergpcmsqpspsr{\hbox{$\erg\cm^{-2}\s^{-1}\sr^{-1}\,$}}
\def\ergpcmcups{\hbox{$\erg\cm^{-3}\s^{-1}\,$}}
\def\ergpcmps{\hbox{$\erg\cm^{-1}\s^{-1}$}}
\def\ergps{\hbox{$\erg\s^{-1}\,$}}
\def\ergpspmp{\hbox{$\erg\s^{-1}\Mpc^{-3}\,$}}
\def\gpcm{\hbox{$\g\cm^{-3}\,$}}
\def\gpcmps{\hbox{$\g\cm^{-3}\s^{-1}\,$}}
\def\gps{\hbox{$\g\s^{-1}\,$}}
\def\Jy{{\rm Jy}}
\def\keVpcmsqpspsr{\hbox{$\keV\cm^{-2}\s^{-1}\sr^{-1}\,$}}
\def\kmps{\hbox{$\km\s^{-1}\,$}}
\def\kmpspmp{\hbox{$\km\s^{-1}\Mpc{-1}\,$}}
\def\Lsunppc{\hbox{$\Lsun\pc^{-3}\,$}}
\def\Msunpc{\hbox{$\Msun\pc^{-3}\,$}}
\def\Msunpkpc{\hbox{$\Msun\kpc^{-1}\,$}}
\def\Msunppc{\hbox{$\Msun\pc^{-3}\,$}}
\def\Msunppcpyr{\hbox{$\Msun\pc^{-3}\yr^{-1}\,$}}
\def\Msunpyr{\hbox{$\Msun\yr^{-1}\,$}}
\def\pcm{\hbox{$\cm^{-3}\,$}}
\def\pcmsq{\hbox{$\cm^{-2}\,$}}
\def\pcmK{\hbox{$\cm^{-3}\K$}}
\def\phpcmsqps{\hbox{$\ph\cm^{-2}\s^{-1}\,$}}
\def\pHz{\hbox{$\Hz^{-1}\,$}}
\def\pmpc{\hbox{$\Mpc^{-1}\,$}}
\def\pmpccu{\hbox{$\Mpc^{-3}\,$}}
\def\ps{\hbox{$\s^{-1}\,$}}
\def\psqcm{\hbox{$\cm^{-2}\,$}}
\def\psr{\hbox{$\sr^{-1}\,$}}
\def\kmpspMpc{\hbox{$\kmps\Mpc^{-1}$}}
% journals
\def\apj{ApJ}
\def\mnras{MNRAS}
\def\nat{Nat}
\def\physrevB{Phys. Rev. B}
\def\araa{ARA\&A}                % "Ann. Rev. Astron. Astrophys."
\def\aap{A\&A}                   % "Astron. Astrophys."
\def\aaps{A\&AS}                 % "Astron. Astrophys. Suppl. Ser."
\def\aj{AJ}                      % "Astron. J."
\def\apjs{ApJS}                  % "Astrophys. J. Suppl. Ser."
\def\pasp{PASP}                  % "Publ. Astron. Soc. Pac."
\def\apjl{ApJ}                   % letter at ApJ
\def\pasj{PASJ}

\voffset=-0.4in

\title[Radiation pressure and absorption in Swift AGN]{Radiation
  pressure and absorption in AGN: results from a complete unbiased  sample
  from Swift} \author[A.C. Fabian, R. V. Vasudevan, R. F. Mushotzky,
L. M. Winter \& C.S. Reynolds] {\parbox[]{6.in} {A.C. Fabian$^1$,
    R. V. Vasudevan$^1$,
    R. F. Mushotzky$^2$, L. M. Winter$^{3,4}$ and C.S. Reynolds$^{4}$\\
    \footnotesize
    $^1$Institute of Astronomy, Madingley Road, Cambridge CB3 0HA\\
    $^{2}$Laboratory for High Energy Astrophysics, NASA/GSFC, Greenbelt, MD 20771, USA\\
    $^{3}$Center for Astrophysics and Space Astronomy, University of Colorado at Boulder, 440 UCB, Boulder, CO 80309-0440, USA\\
    $^{4}$Astronomy Department, University of Maryland, College Park, MD 20742, USA\\
  }}

\maketitle 

\begin{abstract} 
  Outward radiation pressure can exceed the inward gravitational pull
  on gas clouds in the neighbourhood of a luminous Active Galactic
  Nucleus (AGN). This creates a forbidden region for long-lived dusty
  clouds in the observed columnn density -- Eddington fraction
  plane. (The Eddington fraction $\lambda_{\rm Edd}$ is the ratio of
  the bolometric luminosity of an AGN to the Eddington limit for its
  black hole mass.) The Swift/BAT catalogue is the most complete hard
  X-ray selected sample of AGN and has 97 low redshift AGN with
  measured column densities $N_{\rm H}$ and inferred black hole
  masses. Eddington fractions for the sources have been obtained using
  recent bolometric corrections and the sources have been plotted on
  the $N_{\rm H} - \lambda_{\rm Edd}$ plane. Only one source lies in
  the forbidden region and it has a large value of $N_{\rm H}$ due to
  an ionized warm absorber, for which radiation pressure is
  reduced. The effective Eddington limit for the source population
  indicates that the high column density clouds in the more luminous
  objects lie within the inner few pc, where the central black hole
  provides at least half the mass. Our result shows that radiation
  pressure does affect the presence of gas clouds in the inner galaxy
  bulge. We discuss briefly how the $N_{\rm H} - \lambda_{\rm Edd}$
  plane may evolve to higher redshift, when feedback due to radiation
  pressure may have been strong.
  
\end{abstract} 

\section{Introduction}

Evidence for a relatively tight relation between the mass of the
central black hole of a galaxy and the mass or velocity dispersion of
the surrounding stellar bulge (\citealt{1995ARA&A..33..581K};
\citealt{1998AJ....115.2285M}, \citealt{2000ApJ...539L..13G},
\citealt{2001ApJ...555L..79F}; \citealt{2002ApJ...574..740T}) points
to a coupling or feedback of the black hole on the galaxy. The
accretion power produced by the growth of the black hole exceeds the
binding energy of the galaxy by one to two orders of magnitude so some
form of feedback is expected. What is unclear however is how it
works. Is it the energy (\citealt{1998A&A...331L...1S},
\citealt{1998MNRAS.300..817H}) or momentum
(\citealt{1999MNRAS.308L..39F}; \citealt{2002MNRAS.329L..18F};
\citealt{2003ApJ...596L..27K}; \citealt{2005ApJ...618..569M}) supplied
by the central engine that are most effective? Winds, radiation
pressure and jets have all been invoked and it is likely that they all
play a role, although perhaps at different phases and in different
objects.

Direct evidence of feedback is elusive and difficult to obtain. A
quasar at the centre of a galaxy will have little effect on the stars
of the surrounding bulge and only act on the gas. Catching and
recognising an object where the gas is being expelled may not be
easy. Here we provide clear evidence for the effect of radiation
pressure from a central Active Galactic Nucleus (AGN) on surrounding
dusty gas in a hard X-ray, flux-limited sample of AGN.

\section{The Effective Eddington Limit}

\cite{1993ApJ...402..441L}, \cite{1995ApJ...451..510S} and
\cite{2005ApJ...618..569M} have shown that the effective
Eddington limit, when the outward radiation pressure on gas exceeds
the inward gravitational pull, can be much lower for dusty gas than
for ionized dust-free gas, as assumed for the standard derivation.  A
reduction factor of 500 is obtained for an AGN spectral energy
distribution (SED), peaking in the UV and assuming a Galactic
dust-to-gas ratio (\citealt{2008MNRAS.385L..43F}). X-ray ionization
keeps the drift velocity of the dust grains relative to the gas
low. The UV radiation is rapidly used up by a relatively small column
density of gas and dust (equivalent hydrogen column density of $N_{\rm
  H}\sim 10^{21}\psqcm$) and the effective Eddington limit drops off
approximately linearly, reaching unity at column densities where it
goes Compton thick around $N_{\rm H}\sim 10^{24}\psqcm$. This means
that an AGN which has a luminosity one to two orders of magnitude
below that standard Eddington limit may nevertheless exceed the
effective Eddington limit for substantial columns of dusty gas.

The actual interaction of the accelerated face with the rest of an
absorbing cloud will be complicated and likely Rayleigh-Taylor
unstable, with the details depending on magnetic fields, geometry and
on how the force is transmitted to the whole cloud. The net result
however is that long-lived stable clouds are not expected to survive
in a regime where the effective Eddington limit is exceeded. This
manifests as a forbidden region in the column density -- Eddington
fraction plane for AGN (\citealt{2008MNRAS.385L..43F}). By Eddington
fraction, $\lambda_{\rm Edd},$ we mean the ratio of the bolometric
luminosity of the source to the standard Eddington limit. An AGN with
$\lambda_{\rm Edd}=1/500$ is at the Effective Eddington limit for
Galactic abundance dusty gas with $N_{\rm H}\sim 10^{21}\psqcm$. This
is shown in Fig.~1.  This limit has been calculated assuming an
average SED typical for low Eddington ratio AGN (the thick black curve
in Fig.~13 of \citealt{2007MNRAS.381.1235V}); at high Eddington ratios
the substantially different SED shape may marginally reduce the
effective Eddington limit .

The above argument applies only where the black hole dominates the
mass locally.  For outer gas hundreds of pc to kpc from the black
hole, the $\lambda_{\rm Edd}$ boundary increases in proportion to the
enclosed mass of stars. Outer (few kpc) dust lanes in a galaxy bulge
or disk are unlikely to have very high column densities (the total gas
mass at radius $r\kpc$ is otherwise very high, $M_{\rm gas}\approx
10^{10}fN_{\rm 23}r^2\Msun$, where the column density is
$10^{23}N_{23}\psqcm$ and $f$ is the covering fraction), so we suggest
a lower boundary to the forbidden region of $N_{\rm H}\sim 3\times
10^{21}\psqcm$. This agrees with the upper limit to $N_{\rm H}$ of
$10^{21}-10^{22}\psqcm$ found for Seyfert 1--1.2 galaxies by
\cite{Winter:2008az}.

\begin{figure} 
\includegraphics[width=\columnwidth]{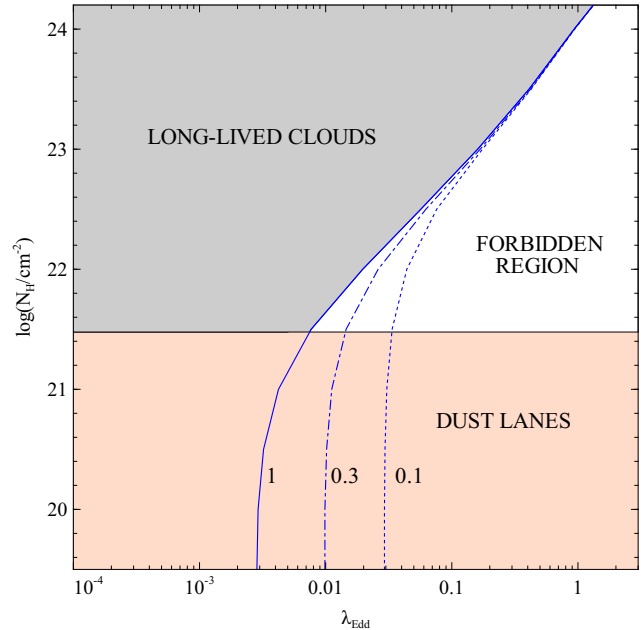}
\caption{ $N_{\rm H} - \lambda_{\rm Edd}$ plane showing the forbidden
  region where dusty clouds close to the black hole see the AGN as
  being effectively above the Eddington limit (blue solid line). Outer
  dust lanes can obscure over much of the plane at low column
  densities (black horizontal line) and long lived absorbing clouds
  can only occur in the upper left region, where the AGN is below the
  effective Eddington limit.  The effective Eddington limit depends on
  the dust grain abundance:  the solid blue line shows the effective
  Eddington limit for a standard ISM grain abundance, and the
  dot-dashed and dotted blue lines are for grain abundances of 0.3 and
  0.1 of ISM abundance respectively.
  \label{eff_edd}}
\end{figure}

\cite{2008MNRAS.385L..43F} plotted several samples on the $N_{\rm H} -
\lambda_{\rm Edd}$ plane. These were a composite low-redshift sample
of 23 objects from {\it Swift} and {\it BeppoSAX}, a sample of 77
sources from the {\it Chandra Deep Field South} and 13 objects from
the {\it Lockman Hole}. The results provided tentative evidence for
sources avoiding the forbidden region. Column densities are derived
from X-ray spectra and Eddington fractions from X-ray luminosity (and
a bolometric correction, \citealt{2007MNRAS.381.1235V}) and
importantly an estimate of the black hole mass. This was mainly done
by combining K-band magnitudes of the AGN host with the $M_{\rm BH} -
K$ correlation of \cite{2003ApJ...589L..21M}. \cite{Winter:2008az}
find that, before application of a bolometric correction, the
distributions of $L_{\rm{X}}/L_{\rm{Edd}}$ for absorbed and unabsorbed
sources are significantly different.  Here we include the bolometric
correction to recover the full Eddington ratio.  We present our
analysis of 97 sources from the 9-month \emph{Swift} BAT catalogue of
AGN, with BH mass estimates (\citealt{2008ApJ...681..113T},
\citealt{Winter:2008az}), which is the most complete and uniform, hard
X-ray flux-limited, sample of AGN. These sources lie above the
Galactic Plane ($|b|>15$~deg) and have 2MASS data for mass
determination.  A hard X-ray sample is important since it will be
relatively insensitive to absorption (provided that the sources are
not too Compton thick).

\section{Results from the Swift/BAT 9-month AGN catalogue}

\begin{figure} 
\includegraphics[width=\columnwidth]{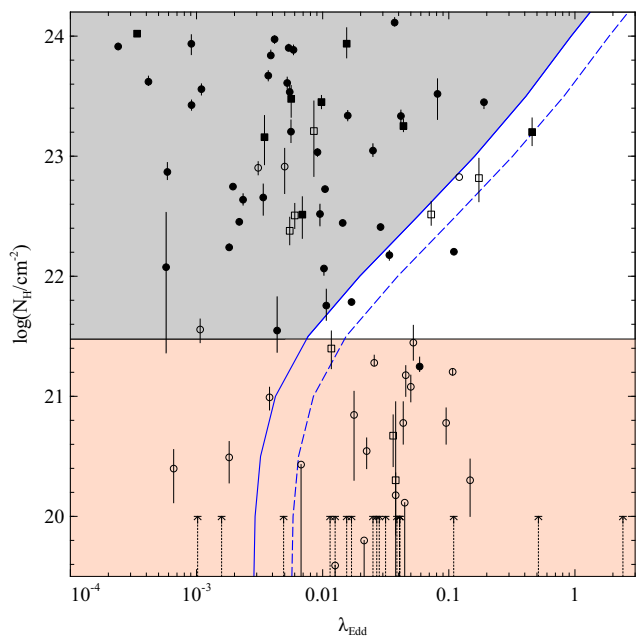}
\caption{Absorbing column density against Eddington ratio for the
  9-month Swift/BAT catalogue of AGN.  Circles represent objects for
  which X-ray fits were taken from the literature by
  \protect\cite{Winter:2008az}, whereas squares represent those for
  which \emph{Swift} XRT spectra were used.  Filled circles and
  squares are those reported to have complex X-ray absorption and
  empty circles and squares are those reported to have simple absorbed
  power-law spectra, in \protect\cite{Winter:2008az}.  The objects
  with arrows pointing upwards in the lower region of the plot did not
  have any column density reported in \protect\cite{Winter:2008az} and
  we show them with a nominal upper limiting column density of
  $10^{20}\rm{cm^{-2}}$. The dashed line shows a factor of 2 increase
  in the effective Eddington limit due to the mass of intervening
  stars.  All other conventions are as given in
  Fig.~\ref{eff_edd} \label{swift_eddabs}}
\end{figure}

In Fig. \ref{swift_eddabs} we present the column density $N_{\rm H}$
against Eddington ratio for the 9-month Swift BAT catalogue of AGN.
The column densities are presented by \cite{Winter:2008az} and were
obtained from fits to X-ray spectra from a variety of X-ray missions
including \emph{ASCA}, \emph{Suzaku}, \emph{Chandra},
\emph{XMM-Newton} and the X-ray telescope (\emph{XRT}) on
\emph{Swift}.  The \emph{XRT} fits were performed by
\cite{Winter:2008az} themselves, whereas the other fits were taken
from the literature\footnote{The source represented in Fig.~2 by an
  empty circle near (0.1, 22.8) is shown after updating its parameters
  using archival \emph{XMM-Newton} data.}.  The Eddington ratios are
calculated using the standard formula $\lambda_{\rm{Edd}}= L_{bol} /
1.38 \times 10^{38} (M_{\rm{BH}}/M_{\rm{\bigodot}})$.
\cite{Winter:2008az} estimate the black hole mass $M_{\rm{BH}}$ using
the $M_{\rm{BH}}-L_{\rm{bulge}}$ relation, using the K-band magnitudes
from the 2MASS All-Sky Survey to provide bulge luminosities.  The
$M_{\rm{BH}}-L_{\rm{bulge}}$ relation from \cite{2006ApJ...637...96N}
is employed by these authors to calculate their BH masses, which are
presented in Table 4 of their work.  We then scale up the 2--10 keV
luminosity to a bolometric luminosity using a hard X-ray bolometric
correction dependent on the X-ray spectral shape.
\cite{2007MNRAS.381.1235V} find that the bolometric correction (and
optical/UV/X-ray SED shape in general) depends on Eddington ratio, and
the study of \cite{2006ApJ...646L..29S} (and
\citealt{2008ApJ...682...81S} subsequently) finds a correlation
between the X-ray photon index, $\Gamma$, and Eddington ratio.  This
latter correlation is not reproduced by \cite{Winter:2008az}, possibly
due to the intrinsic scatter in individual AGN accretion rates and
photon indices, or a systematic deviation of lower luminosity AGN from
the correlation.  However, they do find $\Gamma$ correlates with
$\lambda_{\rm{Edd}}$ in individual objects varying between
observations \citep{2008ApJ...674..686W}, lending credence to the
possibility that spectral shape is linked to accretion rate, albeit in
a non-straightforward fashion.  Based on these findings, we adopt a
bolometric correction of 19 for hard spectrum objects ($\Gamma<1.9$)
and 55 for soft spectrum objects ($\Gamma>1.9$) and thus determine the
Eddington ratios.

We note that the AGN from the 9-month BAT catalogue primarily lie in
the region expected for long-lived absorption or dust lanes.  There
are a few objects which lie above the effective Eddington limit for
dusty gas, but we note that this limit has hitherto been calculated
assuming that the only inward gravitational force of importance is
that from the central black hole.  In reality, there may be a
significant gravitational force from the stars located inward from the
dusty gas clouds of interest.  This would increase the effective
Eddington limit by an amount proportional to the total enclosed mass,
since the increased gravitational force is now able to balance a
proportionately higher outward force due to radiation.

We estimate the importance of this effect as follows.  If we enclose a
mass equal to twice the central BH mass, the effective Eddington limit
is doubled.  Such a limit would allow the majority of the objects to
now lie within the effective Eddington limit, with the exception of
one (MCG-05-23-16).  We can also comment on the constraints that such
a modified effective Eddington limit would place on the location of
the dusty gas clouds.  The studies of \cite{2007A&A...469..125S}
reveal that, in the case of our Galaxy, the enclosed mass doubles at a
scale of a few parsecs from the nucleus.  If nuclear star clusters
often accompany AGN as the study of \cite{2008ApJ...678..116S} seems
to suggest, then an effective Eddington limit that accounts for twice
the BH mass would still comfortably locate the dusty gas clouds close
to the nucleus.

\section{Dust-to-Gas ratio and the Effective Eddington Limit}

We note that the dust-to-gas ratio in AGN may deviate significantly
from the Galactic dust-to-gas ratio
\protect\citep{2001A&A...365...28M}.  The calculations of the
effective Eddington limit in \cite{2008MNRAS.385L..43F} assume the
Galactic value, via the assumption of inter-stellar medium (ISM) grain
abundance in their \textsc{cloudy}
simulations. \protect\cite{2001A&A...365...28M} suggest that the ratio
of optical extinction to X-ray column density $N_{\rm{H}}$ is
systematically lower in AGN, and here we account for this effect by
running further \textsc{cloudy} simulations with dust grain abundances
reduced by factors 0.3 and 0.1 of the standard ISM abundance
(Fig.~\ref{eff_edd}).  The effect of this alteration is signficant at
low $N_{\rm{H}}$, where the effective Eddington limit is
inversely propotional to the dust to gas ratio.  At high column
densities, the effective Eddington limit with the reduced dust-to-gas
ratio converges with that for Galactic dust abundances, since atomic
absorption in the gas dominates.  However, since the change in
effective Eddington limit at low column densities coincides with the
regime where dust lanes could be responsible for the measured
$N_{\rm{H}}$, this modification has little bearing on the shape of the
`forbidden' region in $N_{\rm{H}}-\lambda_{\rm{Edd}}$ space.  We note
that the effective Eddington limit will also depend on the mass of the
black hole, since the peak of the thermal black body in the UV depends
on BH mass, but do not address this further here.  The average SED
used in the \textsc{cloudy} simulations is calculated from eight AGN
SEDs with a range of BH masses \citep{2007MNRAS.381.1235V} so is
probably a representative choice of SED.

%\begin{figure} 
%\includegraphics[width=\columnwidth]{nhvsedd_swiftAGN_dusttogasvaried.eps}
%\caption{  All other conventions are as given in Fig.~\ref{eff_edd}. \label{nhvsedd_grains}}
%\end{figure}

\section{Discussion}

Sources in the most complete sample of hard X-ray selected AGN available
avoid the forbidden region in the $N_{\rm H} - \lambda_{\rm
  Edd}$ plane. This is unlikely to be due to an observational
selection effect since sources there should be brighter. The single
source which lies in the forbidden region is MCG-05-23-016, which has
a complex spectrum with warm absorbers
(e.g. \citealt{2007PASJ...59S.301R}) and so not necessarily absorbed
by dusty gas clouds. Our results provide strong evidence that
radiation pressure from the central AGN is the main agent.

The good agreement between the distribution of sources and our allowed
zones suggests that our assumptions are resonable. The radiation
pressure seems to be transmitted to the whole cloud, or cloud complex,
perhaps by magnetic fields. The Eddington fractions must be accurate
to a factor of a few (\citealt{Winter:2008az} suggest a factor of 3
uncertainty) and therefore the bolometric luminosities also. Our
result (Fig.~2) provides a simple explanation for some of the
otherwise complicated behaviour that has been reported in the
relationship between absorption and luminosity in AGN
(\citealt{2008arXiv0808.0260H} and references therein).

There can be matter out of our line of sight in the low column
density, high Eddington fraction sources but that matter will need to
avoid the forbidden zone. For example it can reside in a Compton thick
torus, as envisaged in many unification models for AGN.

We note that the soft X-ray bright Narrow-Line Seyfert 1 galaxies,
which are generally considered to have high Eddington fractions do not
show up in a hard X-ray selected sample owing to their steep X-ray
spectra \cite{2008MNRAS.389.1360M}. They do not have strong
absorption so would appear at the lower right of the $N_{\rm H} -
\lambda_{\rm Edd}$ plane.

We do not know the evolution of absorbing clouds in these galaxies,
nor the evolution of the central black holes. Thus we cannot say
whether the lowly-absorbed AGN have prevented clouds from accumulating
or have blown away an existing population of clouds. The UV radiation
from the AGN cannot vary greatly on the local dynamical timescale,
$10^4 r_{\rm pc}\sigma_{100}^{-1}\yr$ where $ r_{\rm pc}$ and
$\sigma_{100}$ are the radius in pc and velocity dispersion in units
of $100\kmps$, in order to obtain the agreement we see. Objects can of
course appear in the forbidden zone for timescales much shorter than
this. The absence of such objects argues that the luminosity of these
AGN is either fairly steady or the duty cycle of high luminosity
episodes is small.

What should in future become possible with large samples from XMM,
Chandra and IXO is to study the evolution of the population of
AGN. The distribution of AGN in the $N_{\rm H} - \lambda_{\rm Edd}$
plane seen as a function of redshift should reveal in which direction
they evolve. The mass doubling timescale for the black hole is $\sim
4\times 10^7 \lambda_{\rm Edd}^{-1}\yr,$ assuming an accretion
efficiency of 0.1. Since many of the objects in Fig.~2 have
$\lambda_{\rm Edd}<0.01,$ then they are growing on a timescale $>
4$~Gyr, which is slow. How cold gas accumulates within the host galaxy
bulge and settles near its centre will determine the fuelling of the
AGN and central star formation rates. Cold gas typically resides for
about $10^9\yr$ before forming stars in low $\lambda_{\rm Edd}$
galaxies (e.g. \citealt{1998ARA&A..36..189K}), so star formation
occurs faster than the black hole mass evolves. The rates at which the
mass of the black hole and host galaxy bulge are changing are however
both low.

The evolution of quasars means that high Eddington rate objects with
short doubling times are common around redshifts of two and we can
expect much higher densities of objects near the effective Eddington
limit. The black hole mass is changing faster than gas is turned into
stars and the effect of radiation pressure on the expulsion of gas can
have dramatic effects on the mass reservoir for growing stars as well
as the black hole (see \citealt{1999MNRAS.308L..39F},
\citealt{2005ApJ...618..569M} and \citealt{2008MNRAS.385L..43F} for
possible scenarios). The locus of the effective Eddington limit in the
$N_{H}-\lambda_{\rm{Edd}}$ plane specifies the radius of the absorbing
gas within the host galaxy and thus the masses of gas
involved. Although we cannot deduce the evolution of an individual
object from its position in the plane, how the population evolves may
be attainable.

Our results show direct evidence for a coupling of the radiation from
AGN and the surrounding gas.

%\section{Summary}

\section{Acknowledgments}

ACF thanks The Royal Society for support, RVV acknowledges support from
the UK Science and Technology Funding Council (STFC). We thank the BAT
Team for their work which has made our study possible. 

\bibliographystyle{mnras} %% MNRAS style bibliography conventions
\bibliography{swift_eddabs}

\end{document}